\documentclass[aps,twocolumn,nofootinbib,showpacs,floatfix]{revtex4}
\usepackage{graphics} 
\usepackage{amssymb,amsmath}
\usepackage{amsmath}
\usepackage{graphicx}
\tighten 

\newcommand{\be}{\begin{equation}} 
\newcommand{\ee}{\end{equation}} 
\newcommand{\bea}{\begin{eqnarray}} 
\newcommand{\eea}{\end{eqnarray}}  
\newcommand{\bean}{\begin{eqnarray*}} 
\newcommand{\eean}{\end{eqnarray*}}

\def\lsim{\raise 0.4ex\hbox{$<$}\kern -0.8em\lower 0.62ex\hbox{$\sim$}} 
\def\gsim{\raise 0.4ex\hbox{$>$}\kern -0.7em\lower 0.62ex\hbox{$\sim$}}

\newcommand{\bse}{\begin{subequations}}
\newcommand{\ese}{\end{subequations}}

\newcommand{\cd}{{\langle n(r) \rangle_p}}

\def\spose#1{\hbox to 0pt{#1\hss}}      
\def\ltapprox{\mathrel{\spose{\lower 3pt\hbox{$\mathchar"218$}}      
\raise 2.0pt\hbox{$\mathchar"13C$}}}      
\def\gtapprox{\mathrel{\spose{\lower 3pt\hbox{$\mathchar"218$}}      
\raise 2.0pt\hbox{$\mathchar"13E$}}}      
\def\inapprox{\mathrel{\spose{\lower 3pt\hbox{$\mathchar"218$}}      
\raise 2.0pt\hbox{$\mathchar"232$}}}

\begin{document} 
\title{Gravitational  Structure Formation, the Cosmological Problem and Statistical Physics}
\author{Luciano Pietronero}   
\affiliation{Dipartimento di Fisica, 
Universit\`a ``La Sapienza'', P.le A. Moro 2, I-00185 Rome, Italy and ISC-CNR Via dei Taurini, 19 00185 Rome Italy}
\author{Francesco Sylos Labini}   
\affiliation{ ``E. Fermi'' Center, Via Panisperna 89 A, Compendio del 
Viminale, 00184 - Rome, Italy and ISC-CNR Via dei Taurini, 19 00185
Rome Italy}
\begin{abstract}   
\begin{center}    
{\large\bf Abstract}   
\end{center}    
Models of structure formation in the universe postulate that matter
distributions observed today in galaxy catalogs arise, through a
complex non-linear dynamics, by gravitational evolution from a very
uniform initial state. Dark matter plays the central role of providing
the primordial density seeds which will govern the dynamics of
structure formation. We critically examine the role of cosmological
dark matter by considering three different and related issues: Basic
statistical properties of theoretical initial density fields, several
elements of the gravitational many-body dynamics and key correlation
features of the observed galaxy distributions are discussed, stressing
some useful analogies with known systems in modern statistical
physics. 
\end{abstract}    
\maketitle   
\date{today}  

\twocolumngrid   

\section{Introduction}

The large distribution of matter in the universe as traced by galaxy
structures shows a complex irregular pattern, characterized by
clusters of galaxies which are organized in filaments around large
voids. In the framework of standard cosmological models these
structures arise from non-linear dynamical evolutions in which
Newtonian gravity plays the essential role
(e.g., \cite{pee_80}). The topic of structure formation in the
expanding universe is clearly extremely large and we focus here on
three different and related issues, where the methods and concepts of
statistical physics find a fruitful applications
\cite{book}.

\section{Primordial density fields} 

The first issue we consider concerns the statistical properties of
initial matter density fields in standard cosmological models: as we
discuss below, in the framework of Friedmann-Robertson-Walker (FRW)
\cite{pee_80} models of cosmological expansion, there are important
constraints which have to be satisfied by primordial matter
fluctuation fields and which are common to all models independently 
on the nature of dark matter \cite{glass}. Some very interesting
analogies with glassy systems (e.g. one component plasma) will be
outlined and the crucial observational tests of standard models of
cosmological structure formation will be discussed.

Dark matter plays the major role in the problem of structure formation
in standard cosmological models: large-scale structures we observe
today must have formed from the effects of gravity acting on small
amplitude seed fluctuations in the original distribution of dark
matter from the Big Bang. The problem of structure formation in the
universe is approximately Newtonian but embedded in an expanding
universe, whose dynamics is described by General Relativity.  The role
and amount of dark matter is determined to make compatible different
types of astronomical observations with the FRW models.  Without
entering into the details of the various reasons why dark matter is
fundamental in the cosmological context it is worth noticing here that
the most ``popular'' model, which nowadays is the so-called
$\Lambda$-Cold Dark Matter (or $\Lambda$CDM), postulates the structure
of mass and energy in a Universe as $5\%$ ordinary baryonic matter,
$25\%$ CDM of non-baryonic form (which has not been yet detected in
laboratories on Earth), and $70\%$ dark energy, which would be a
uniform component of energy repelling to gravity and pushing the
Universe apart faster.  While this latter energy component is
essentially relevant only for the rate of expansion in FRW models, the
CDM would be important for structure formation and ordinary baryonic
matter does not make any relevant dynamical effect and it just follows
the distribution of CDM.

Primordial density fluctuations have imprinted themselves on the
patterns of radiation also, and those variations should be detectable
in the Cosmic Microwave Background Radiation (CMBR).  Three decades of
observations have revealed fluctuations in the CMBR of amplitude of
order $10^{-5}$ \cite{wmap}. It is in fact to make these measurements
compatible with observed structures that it is necessary to introduce
non-baryonic dark matter which interact with photons only
gravitationally, and thus in a much weaker manner than ordinary
baryonic matter.  Thus in these models dark matter plays the dominant
role of providing density fluctuation seeds which, from the one hand
are compatible with observations of the CMBR and from the other hand
they are large enough to allow the formation, through a complex non
linear dynamics, of galaxy structures we observe today.

A simple calculation may give the order of magnitude of the initial
amplitude of fluctuations. In the linear regime small amplitude
fluctuations grow inversely proportional to the redshift $z$ in the
expanding universe.  Thus given that the CMBR is at $z=10^3$
(i.e. early cosmological times), in order to have density fluctuations
$\delta\rho/\rho$ of order one today (i.e. $z=0$) on a certain scale
(and larger than the average on smaller scales to make non-linear
structures) one should have $\delta\rho/\rho
\approx 10^{-3}$ at $z=10^3$ at the corresponding scale. However there
is a factor $10^2$ of difference given that $\delta T / T
\approx 10^{-5}$ and $\delta\rho/\rho \approx \delta T / T$ 
for ordinary baryonic matter. As mentioned, CDM interacts weakly with
radiation making possible to have $\delta\rho/\rho \approx 10^{-3}$ at
$z=10^3$ and in the same time $\delta T / T \approx 10^{-5}$. This
would not be possible with ordinary baryonic matter.  Therefore
properties of dark matter in this cosmological context are defined to
allow structure formation today.

From the above discussion it seems that much freedom is left for the
choice of dark matter, its physical properties and its statistical
distribution, unless it will once be directly observed. However there
is an important constraint which must be valid for any kind of initial
matter density fluctuation field in the framework of FRW models.  This
must be imprinted in the CMBR, a relict of the high energy process
occurred in the early universe according to standard models.

The most prominent feature of theoretical models of the initial
conditions derived from inflationary mechanisms is that matter density
field presents on large scale super-homogeneous features~\cite{glass}.
This means the following. If one considers the paradigm of uniform
distributions, the Poisson process where particles are placed
completely randomly in space, the mass fluctuations in a sphere of
radius $R$ growths as $R^3$, i.e. like the volume of the sphere. A
super-homogeneous distribution is a system where the average density
is well defined (i.e. it is uniform) and where fluctuations in a
sphere grow slower than in the Poisson case, e.g. like $R^2$: in this
case there are the so-called surface fluctuations to differentiate
them from Poisson-like volume fluctuations.  (Note that a uniform
system with positive correlations present fluctuations which grow
faster than Poisson.)  For example a perfect cubic lattice of particle
is a super-homogeneous system. An example of a well known system in
statistical physics systems of this kind is the one component
plasma~\cite{lebo} which is characterized by a dynamics which at
thermal equilibrium gives rise to such configurations.  In the
cosmological context inflationary models predict a spectrum of
fluctuations of this type.

The reason for this peculiar behavior of primordial density
fluctuations is the following.  In a FRW cosmology there is a
fundamental characteristic length scale, the horizon scale
$R_H(t)$. It is simply the distance light can travel from the Big Bang
singularity $t=0$ until any given time $t$ in the evolution of the
Universe, and it grows linearly with time.  The Harrison-Zeldovich
(H-Z) criterion states that the normalized mass variance at the
horizon scale is constant: this can be expressed more conveniently in
terms of the power spectrum (PS) of density fluctuations
\cite{glass}
$P(\vec{k})=\left<|\delta_\rho(\vec{k})|^2\right>$ 
where $\delta_\rho(\vec{k})$ is the Fourier Transform of the
normalized fluctuation field $(\rho(\vec{r})-\rho_0)/\rho_0$, being
$\rho_0$ the average density. It is possible to show that
the H-Z-criterion is equivalent to assume $P(k) \sim k$: in this
situation matter distribution present surface fluctuations 
\cite{glass}.

In order to illustrate more clearly the physical implications of this
condition, one may consider gravitational potential fluctuations
$\delta\phi(\vec{r})$ which are linked to the density fluctuations
$\delta\rho(\vec{r})$ via the gravitational Poisson equation:
%
$\nabla^2\delta\phi(\vec{r})=-4\pi G \delta\rho(\vec{r})\,. 
$
%
From this, transformed to Fourier space, it follows that the PS of the
potential $P_{\phi}(k)=\left<|\delta\hat\phi(\vec{k})|^2\right>$ is
related to the density PS $P(k)$ as: $P_{\phi}(k)\sim
\frac{P(k)}{k^4}\,$. The H-Z condition corresponds therefore to
$P_{\phi}(k) \propto k^{-3}$ so that gravitational potential
fluctuations become constant as a function of scale.

The H-Z condition is a consistency constraint in the framework of FRW
cosmology. In fact the FRW is a cosmological solution for a
homogeneous Universe, about which fluctuations represent an
inhomogeneous perturbation: if density fluctuations obey to a
different condition than the H-Z criterion, then the FRW
description will always break down in the past or future, as the
amplitude of the perturbations become arbitrarily large or small.  
For
this reason the super-homogeneous nature of primordial density field
is a fundamental property independently on the nature of dark
matter. We note that this is a very strong condition to impose, and it
excludes even Poisson processes ($P(k)=$const. for small $k$)
\cite{glass}.

This is the behavior that one would like to detect in the data in
order to confirm inflationary models. Up to now this search has been
done through the analysis of the galaxy PS which has to go
correspondingly as $P(k) \sim k$ at small $k$ (large scales). No
observational test of this behavior has been provided yet.


\section{The gravitational many-body problem}

As mentioned, the standard model of the formation of large scale
structure of the universe is based on the gravitational growth of
small initial density fluctuations in a homogeneous and isotropic
medium (e.g., \cite{pee_80}). In the CDM model particles interact only
gravitationally and they are cold, i.e. with very small initial
velocity dispersion. This situation allows to model this system with a
collision-less Boltzmann equation and, for sufficiently large scales,
pressure-less fluid equations. Then it is possible to solve in a
perturbative way, for small density fluctuations, these fluid
equations (for a review see e.g. \cite{pee_80}). However
this treatment is inapplicable in the strong non-linear regime. Then,
the most widely used tool to study gravitational clustering in the
various regimes is by means of N-body simulations (NBS) which are
based on the computation of particle gravitational dynamics in an
expanding universe.

One may consider an infinite periodic system, i.e. a finite system
with periodic boundary condition. Despite the simplicity of the
system, in which dynamics is Newtonian at all but the smallest scales,
the analytic understanding of this crucial problem is limited to the
regime of very small fluctuations where a linear analysis can be
performed. As mentioned, the problem is Newtonian but the equations of
motion are modified because of the expanding background. It is
possible to consider some simplified cases where the expansion is not
included and then study the differences introduced by space expansion.

An additional important point should be stressed: for numerical
reasons, the cosmological density field must be discretized into
``macro-particles'' interacting gravitationally which are tens of
order of magnitudes heavier than the (elementary) CDM particles due to
computer limitations. This procedure introduces discreteness at a much
larger scale than the discreteness inherent to the CDM particles. By
discreteness we mean statistical and dynamical effects which are not
described by the self-gravitating fluid approximation. The
discreteness has different manifestations in the evolution of the
system (see e.g. \cite{grav1} and references therein).  In this
context it is necessary to consider the issue of the physical role of
discrete fluctuations in the dynamics, which go beyond a description
where particles play the role of collision-less fluid elements and the
evolution can be understood in terms of a self-gravitating fluid.  

In order to study the full gravitational many-body problem, we have
considered a very simple initial particle distribution represented by
a slightly perturbed simple cubic lattice with zero initial velocities
\cite{letterlinear}. A perfect cubic lattice is an unstable
equilibrium configuration for gravitational dynamics, being the force
on each particle equal to zero. A slightly perturbed lattice
represents instead a situation where the force on each particle is
small, and linearly proportional to the average root mean square
displacement of any particle from its lattice position. When the
system is evolved for long enough times it creates complex non-linear
structures. While the full understanding of this clustering dynamics
is not currently available, some steps have been done for what
concerns the early times evolution of the system
\cite{letterlinear,bsl04}.

In this context an analogy with the dynamics of the Coulomb lattice
(or Wigner crystal) helps to develop an analytical approach to the
early time evolution of a gravitational infinite lattice of point mass
particles, slightly perturbed about the equilibrium configuration.
Apart a change in the sign of the force (in the Coulomb lattice case
it is repulsive) the equation of motion is identical in the
gravitational and Coulomb cases \cite{letterlinear}.  This allows to
quantitatively characterize and understand, in the same approximation,
the deviation of a finite number of particle system from the evolution
of a self-gravitating fluid. This is relevant for the problem of
cosmological structure formations, where the fluid approximation is
usually used to model a system of large number of elementary dark
matter particles while the simulations used to study numerically the
problem employ a relative small number of particles
\cite{letterlinear,bsl04}. 

This analogy is extremely useful to treat the particular problem of
the evolution of a shuffled lattice at early times, i.e. before
nearest particles collide. In this case one may show that there is a
rich structure of non-fluid modes, including some modes which grow
faster than the fluid one \cite{letterlinear}.  Then a study of the
Poisson distribution \cite{bsl04} is very useful to treat the
formation of the first power-law correlated structures. It is then
surprising that the early-times power-law correlation function remains
invariant during the subsequent time evolution, when structures are
formed by many particles. A study of the late times dynamics is
currently under consideration.

\section{Large Scale Structures of the universe}

The third topic we will briefly discuss is represented by the
statistical properties of the matter distribution observed today
through the study of three-dimensional galaxy catalogs \cite{sdss}.
Galaxy correlation properties seem to be similar to those of a fractal
object \cite{joyceetal2005}.  This distribution represents the
observational test for any theory of cosmological structure formation.

\begin{figure}
\includegraphics*[width=0.5\textwidth]{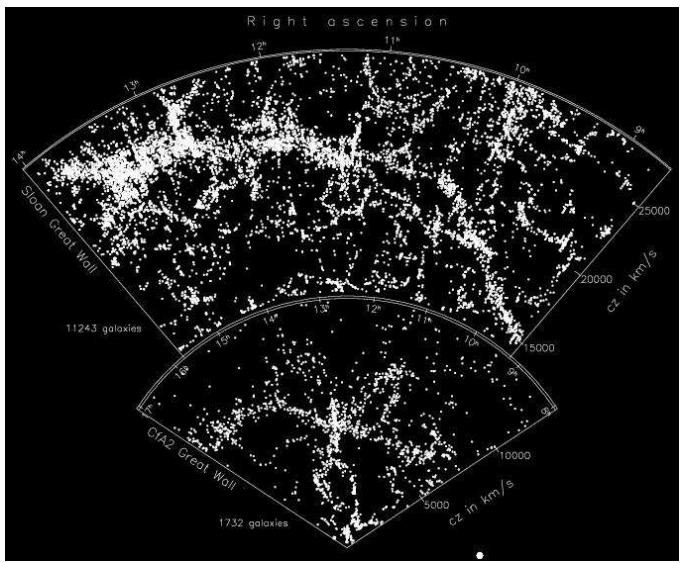}
\caption{ 
Progress in redshift surveys: it is reported the ``slice of the
universe'' from the CfA2 redshift survey \cite{delapp} (lower part)
and the new SDSS data \cite{gott} (upper part).  This cone diagram
represents the reconstruction of a thin slice observed from the Earth
which is in the bottom.  The CfA2 slice has an depth of 150 Mpc/h,
while the SDSS slice has a depth of 300 Mpc/h.  The ``Great Wall'' in
the CfA2 slice and the new ``Sloan Great Wall'' in the SDSS slice are
the dominant structures in these maps and they are clearly
recognizable.  For comparison we also show a small circle of size of 5
Mpc/h (bottom of the figure), the typical clustering length separating
the regime of large and small fluctuations according to the standard
analysis. (Elaboration from
\cite{book}.)
\label{sdss}}
\end{figure}

In the past twenty years observations have provided several three
dimensional maps of galaxy distribution, from which there is a growing
evidence of large scale structures.  This important discovery has been
possible thanks to the advent of large redshift surveys: angular
galaxy catalogs, considered in the past, are in fact essentially smooth and
structure-less. Figure \ref{sdss} shows a slice of the Center for
Astrophysics galaxy catalog (CfA2), which was completed in the early
nineties \cite{delapp}, and a slice derived from the recent
observations of the Sloan Digital Sky Survey (SDSS)
project~\cite{sdss}.  In the CfA2 catalog, which was one of the first
maps surveying the local universe, it has been discovered the giant
``Great Wall'' a filament linking several groups and clusters of
galaxies of extension of about $200$ Mpc/h
\footnote{The typical mean separation between nearest galaxies is of
about 0.1 Mpc. By local universe one means scales in the range
$[1,100]$ Mpc/h, where space geometry is basically Euclidean and
dynamics is Newtonian, i.e. effects of General Relativity are
negligible. On larger scales instead, one has to consider that
relativistic corrections start play a role for the determination of
the space geometry and dynamics.  The size of the universe, according
to standard cosmological models is about 5000 Mpc/h, where 1 Mpc
$\simeq 3 \times 10^{22}$ m; distances are given in units of h, a
parameter which is in the range [0.5,0.75] reflecting the incertitude
in the value of the Hubble constant ($H$=100 h km/sec/Mpc) used to
convert redshift $z$ into distances $d\approx c/H z$ (where $c$ is the
velocity of light).} and whose size is limited by the sample
boundaries.  Recently the SDSS project has reveled the existence of
structures larger than the Great Wall, and in particular in
Fig.\ref{sdss} one may notice the so-called ``Sloan Great Wall'' which
is almost double longer than the Great Wall. Nowadays this is the most
extended structure ever observed, covering about 400 Mpc/h, and whose
size is again limited by the boundaries of the sample~\cite{gott}.

The search for the ``maximum'' size of galaxy structures and voids,
beyond which the distribution becomes essentially smooth, is still an
open problem. Instead the fact that galaxy structures are strongly
irregular and form complex patterns has become a well-established
fact.  From the theoretical point of view the understating of the
statistical characterization of these structures represents the key
element to be considered by a physical theory dealing with their
formation.  The primary questions that such a situation rises are
therefore: (i) which is the nature of galaxy structures and (ii) which
is the maximum size of structures ?  A number of statistical concepts
can be used to answer to these questions: in general one wants to
characterize $n$-point correlation properties which are able to capture
the main elements of points distributions~\cite{book}.

Recently a team of the SDSS collaboration~\cite{hoggetal2004} has
measured the conditional density $\cd$ (which gives the average
density at distance $r$ from an occupied point) as a function of scale
in a sample of the SDSS survey which covers, to date, the largest
volume of space ever considered for such an analysis with a very
robust statistics and precise photometric calibration
(Fig.\ref{sdss}).  They found that: (i) There is clearly a ``fractal
regime'' where $\cd \sim r^{D-3}$ with a dimension $D
\approx 2$, which appears to terminate at somewhere between 20 and 30
Mpc/h --- this behavior agrees very well with what we found at the
scales we could probe properly (i.e. by making the full volume
average) with the samples at our disposal a few years
ago~\cite{slmp98} and recently with the new 2dF sample \cite{niko}
(see discussion in \cite{joyceetal2005}). (ii) The data show then a
slow transition to homogeneity in the range $30 < r < 70$ Mpc/h, where
a flattening of the conditional density seems to occur for scales
larger than $\lambda_0 \approx 70$ Mpc/h, a scale comparable with the
sample size precisely where its statistical validity becomes weaker.
Note that often in the past, samples have shown finite size effects
which produced this type of behavior, which was then eliminated by
deeper samples (see e.g. \cite{bt05}).  For example such a high value
of $\lambda_0$ implies that {\it all} previous determinations of the
characteristic clustering length $r_0$ are biased by finite size
effects ~\cite{joyceetal2005}.  In fact the estimated $r_0$ has grown
of about a factor 3 from $5$ Mpc/h to about $13$ Mpc/h in the most
recent data~\cite{zehavi,joyceetal2005}.

Whether the latest measurements will remain stable in future larger
samples is a key issue to be determined, and this is directly related
to the reality of the flattening at 70 Mpc/h: This will be clarified
soon, as the volume surveyed by the SDSS will increase rapidly in the
near future.


\section{Conclusions} 

Statistical properties of primordial density fields show interesting
analogies with systems in statistical physics, like the one-component
plasma, whose main characteristic is the ordered, or
super-homogeneous, nature.  In the FRW models the super-homogeneous
(or Harrison-Zeldovich) condition arises as a kind of consistency
constraint: other, more inhomogeneous, stochastic fluctuations, like
the uncorrelated Poisson case, will always break down in the FRW
models in the past or future as the amplitude of perturbations in the
gravitational potential may become arbitrarily large.  We discussed
that the observational detection of the super-homogeneous character of
the matter density field, through the observation of galaxy
distribution or of the CMBR anisotropies, is still lacking.  On the
other hand the main feature of galaxy two-point correlation function
is represented by its power-law character in the strongly non-linear
region. We stressed that a clear crossover to homogeneity is also not
well established in the data, and thus the transition from the highly
clustered phase to the highly uniform (super-homogeneous) one is the
main observational tests for theories of the early universe.  For
galaxies this should be evidenced as a negative correlation function
behaving as $-r^{-4}$ (corresponding to $P(k) \sim k$) at large
scales. In this way one may have some constraints on the large number
of free parameters which characterize cosmological dark matter, the
main source for the seeds of structure formation in the universe
according to standard models.

The theoretical understating of non-linear structure formation of a
self-gravitating infinite particle distribution is a fascinating
problem and many questions are still open.  We discussed the fact that
at early times, starting from an instable equilibrium configuration as
a simple cubic lattice of point mass particles, it is possible to
develop a stringent analogy with the Coulomb lattice dynamics. In this
way it is possible to characterize and understand the deviation of a
finite number of particle system from the evolution of a
self-gravitating fluid. That is, it is possible to quantify the
effects of discreteness and their role in the formation of non-linear
structures. In this respect the main problem of cosmological
simulations is that, because of the discretization used for numerical
limitations, they could be affected by discrete effects, as a particle
in the simulations has a mass which is many orders of magnitude larger
than the elementary dark matter particles one would like to
simulate. The understanding of the full time evolution, and of the
creation of non-linear structures made by many particles is still the
main open problem in this context.

\begin{acknowledgments}
We thank T. Baertschiger, A. Gabrielli, M. Joyce and B. Marcos for
fruitful collaborations.
\end{acknowledgments}

\onecolumngrid

{}

\end{document}